# Information Retrieval (IR) through Semantic Web (SW): An Overview

**Gagandeep Singh[1], Vishal Jain[2]**

*[1]B.Tech (CSE) VI Sem, GuruTegh Bahadur Institute of Technology, GGS Indraprastha University, Delhi*
*[2]Research Scholar, Computer Science and Engineering Department, Lingaya's University, Faridabad*
*[1]gagan.narula87@gmail.com,[2]vishaljain83@ymail.com*

***ABSTRACT***

*A large amount of data is present on the web. It contains huge number of web pages and to find suitable information from them is very cumbersome task. There is need to organize data in formal manner so that user can easily access and use them. To retrieve information from documents, we have many Information Retrieval (IR) techniques. Current IR techniques are not so advanced that they can be able to exploit semantic knowledge within documents and give precise results. IR technology is major factor responsible for handling annotations in Semantic Web (SW) languages and in the present paper knowledgeable representation languages used for retrieving information are discussed.*

## 1. INTRODUCTION

We view the future web as combination of text documents as well as Semantic markup. Semantic Web (SW) uses Semantic Web documents (SWD's) that must be combined with Web based Indexing. Current IR techniques are not so intelligent that they are able to produce semantic relations between documents. Extracting information manually with the help of Extensible Markup Language (XML) and string matching techniques like Rabin Karp matcher has not proven successful. To use these techniques normal user has to be aware of all these tools.

So, keeping this in mind we have moved to concept of Ontology in Semantic Web. It represents various languages that are used for building semantic web (SW) and increase accuracy.

## 2. SEMANTIC WEB (SW)

In spite of many efforts by researchers and developers, SW has remained a future concept or technology. It is not practiced presently.

There are few reasons for this which is listed below:

(a) Complete Semantic Web (CSW) has not been developed yet and the parts that have been developed are so poor that they can't be used in real world.

(b) No optimal software or hardware is provided.

"SW is not technology, it is philosophy" [1]. It is defined as collection of information linked in a way so that they can be easily processed by machines. From this statement, we conclude that SW is information in machine form. It is also called *Global Information Mesh (GIM)* [2]. This is also known as framework for expressing information.

### *II.1 PRINCIPLE OF SW*

Both, Semantic Web (SW) and World Wide Web (www) are entirely different from each other. SW is machine understandable while www is machine readable. Current SW languages like Resource Description Framework (RDF) do not work with www.The following. Fig. 1 describes the structure of semantic web.

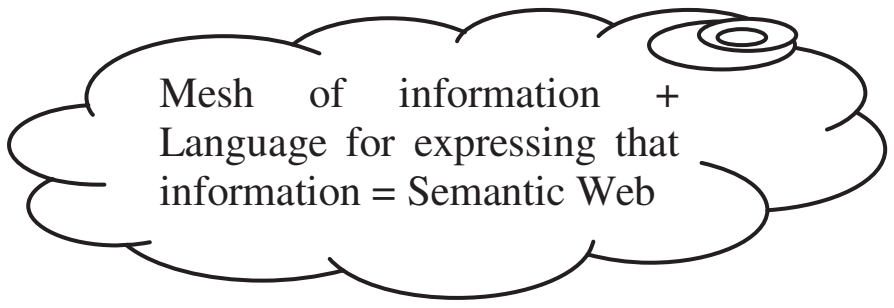

**Fig. 1: "Semantic Web principles"**

### *II.2 SEMANTIC WEB ARCHITECTURE*

Architecture consists of following parts:

- Uniform Resource Identifier (URI) and UNICODE: - Semantic Web contains URI's to represent data in triples based structures with the help of syntaxes designed for particular task.
- UNICODE supports intellectual text of style.
- RDF and rdf schema: - RDF processes metadata and provides interoperation to work together between applications that exchange machine understandable information on web while Rdf schema is RDF vocabulary description language and represents relationship between group of resources. There is RDF model (Figure2) described below representing properties and their values.

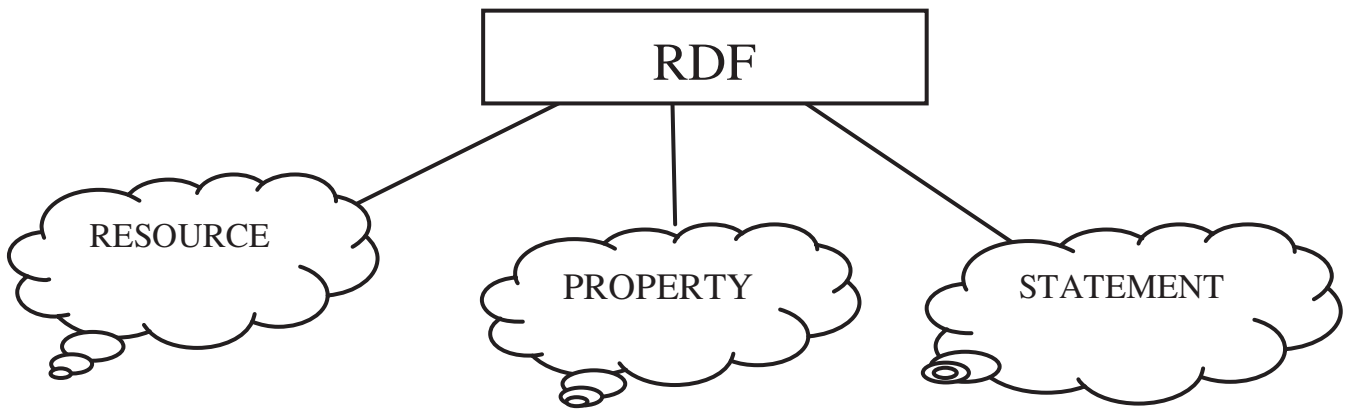

**Fig. 2: "RDF Model"**

*Resource* may be web pages or individual elements of XML document. Resource with its name is called *Property*. *Statement* is combination of Resource and Property and its value.

E.g.:- Gagan plays football. In this sentence, Gagan is object, plays are his property and football is resource.

Football plays Gagan →

- **Ontology:** - Ontology is abbreviated as FESC which means Formal, Explicit, specification of Shared Conceptualization [3].

Formal specifies that it should be machine understandable. Explicit defines the type of constraints used in the model. Shared defines that ontology is not for individual, rather it is for group. Conceptualization means model of some phenomenon that identifies relevant concept of that phenomenon.

**Inference: -** This is defined as producing new data from existing one or to reach some conclusion, e.g. Adios is a French word which is replaced by Good bye that is understandable by user.

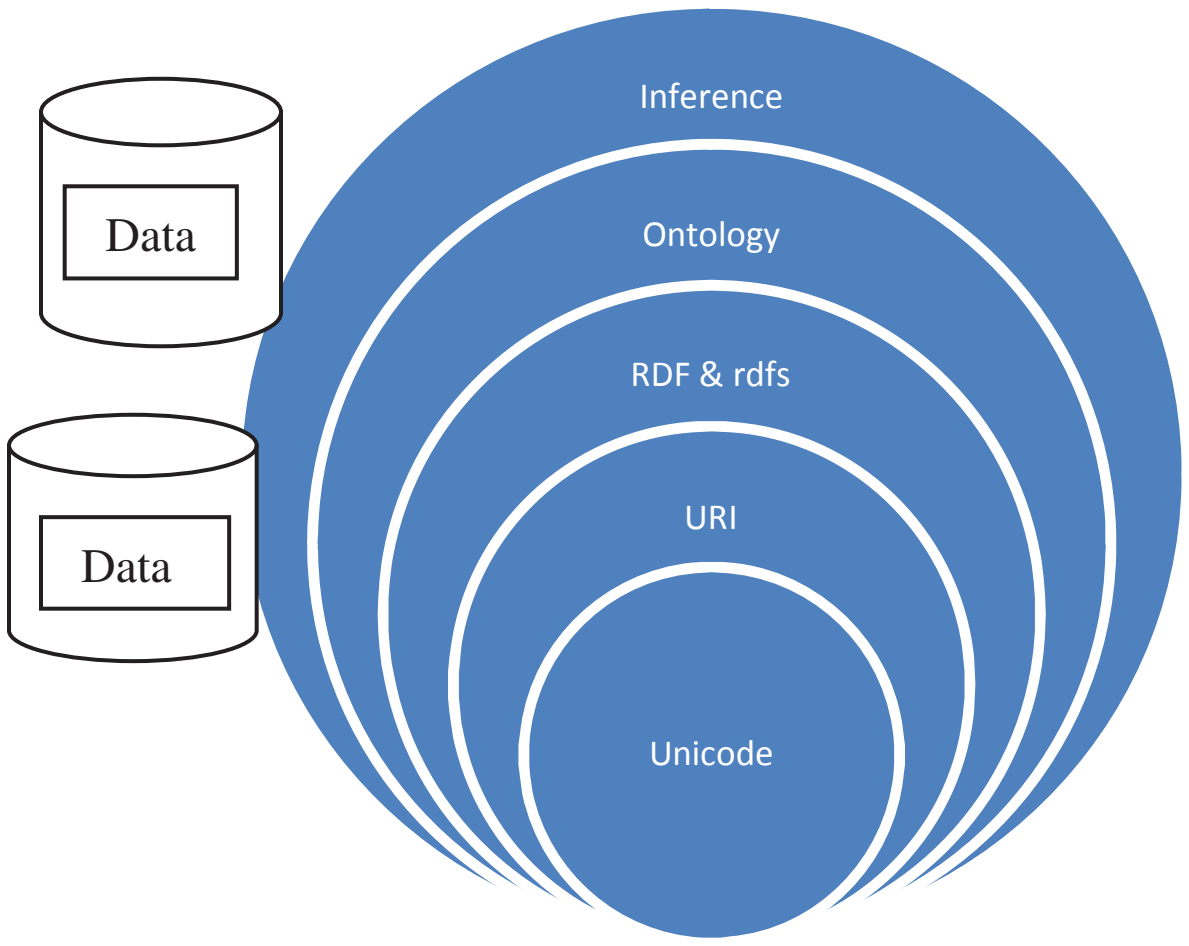

**[4]Fig. 3:"SW Architecture"**

## 3. INFORMATION RETRIEVAL (IR)

IR involves identifying and extracting relevant pages containing that specific information according to predefined guidelines. There are many IR techniques for extracting keywords like NLP based extraction techniques which are used to search for simple keywords. Then we have Aero Text system for text extraction of key phrases from text documents.

### *III.1 IR PROCESS and ARCHITECTURE*

How we retrieve information? The answer to this question explained below.

Background knowledge is stored in form of ontology that can be used at any step. As we have ranked list of documents, they are indexed to form document in represented way.These documents produce ranked results which are given to admin. Admin solves user query which leads to transformation of user query.

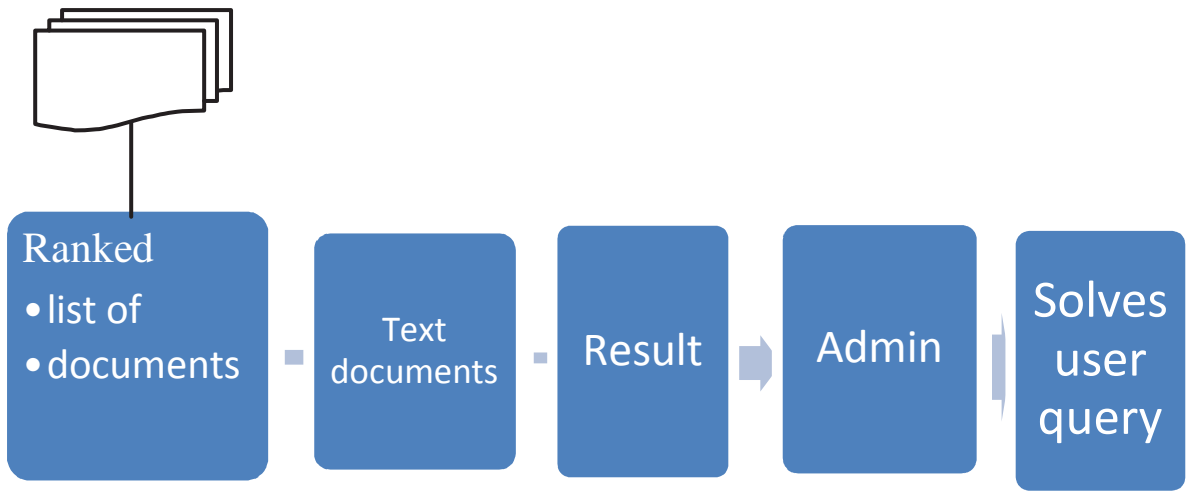

**Fig. 4: "Retrieval of Information"**

### *III.2 ARCHITECTURE*

It is based on ONTOLOGY BASED MODEL [5] that represents the content of resource from given ontology.

It has following parts:

- OMC (Ontology Manager Component):- This is used by Indexer, Search Engine and GUI.
- INDEXER: This indexes documents and creates metadata.
- SEARCH ENGINE
- GUI supports user in query formation.

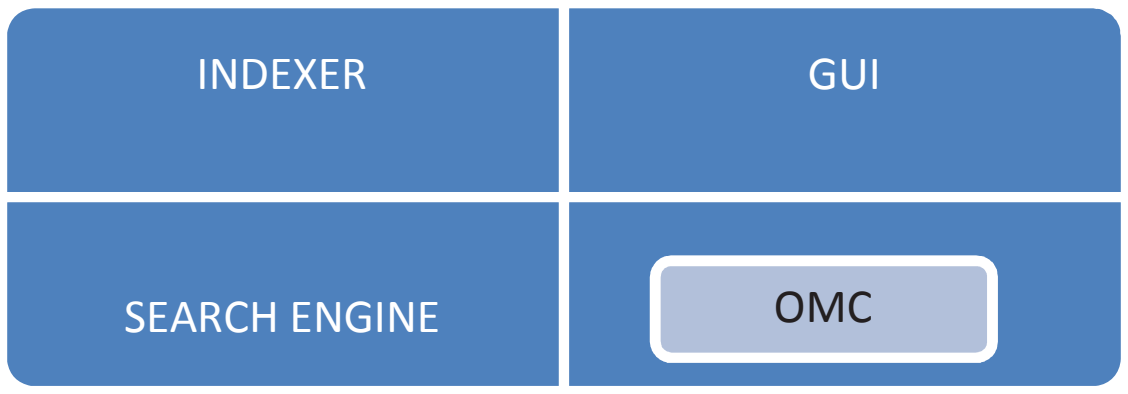

**Fig. 5: "Architecture"**

## 4. HIR (HYBRID INFORMATION RETRIEVAL)

Since the standard IR approaches used can create differences among the documents by adding additional features. So, to avoid differences and make complete document we have used HIR approach and documents developed using HIR, called HYBRID documents.

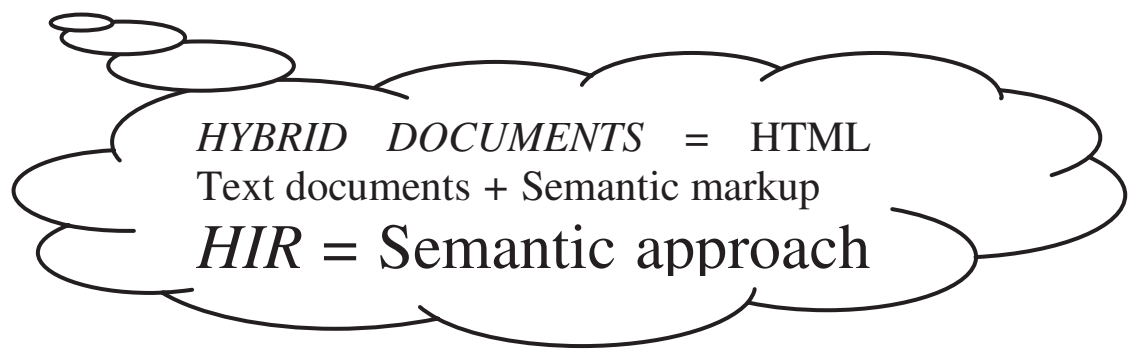

**Fig. 6: "Hybrid Documents"**

### *A. Components of HIR*

**Table 1: "Components of HIR"**

| Standard Text IR | Semantic IR |
|---|---|
| It contains Vector Space Model, Indexing and Markup similarity | It contains Inference, Ontology mapping, Markup relations. |

***Markup/ Text relationship:*** It defines information about how many times the Semantic markup occurs in data. It converts text query into Semantic markup.

***Markup Similarity:*** It allows ranking results together with text documents.

## 5. PROTOTYPE SYSTEMS

After using several approaches for retrieving information from documents, we have developed three prototype systems that make use of knowledgeable representation languages for solving queries. They are (i) OWLIR, (ii) SWANGLER and (iii) SWOOGLE. These are discussed below.

### *A. OWLIR*

***Problem*: -** When we want to retrieve text and semantic documents, there is not surety that we get relevant ones and traditional text search engine uses text only. Then what is the way to retrieve SW documents.

***Solution*: -** OWLIR

***Analysis*: -** OWLIR is an acronym for *Ontology Web Language and Information Retrieval*. It is a system for retrieval of text as well as semantic markup documents in languages like DAML+OIL, RDF and OWL.OWLIR follows three processes which are used to access both semantic web pages and text documents.

- IR: - gathering information about documents for query.
- Q & A: Ask simple questions and answers.
- Complex Q & A

**OWLIR** works with two retrieval engines- HAIRCUT and WONDIR.

**HAIRCUT: -** It is abbreviated as Hopkins Automated Information Retrieval (HAIR) for Combining Unstructured Text (CUT). It is used for specifying required query terms. This is also language modeling approach to find similarity between documents.

**WONDIR: -** It is abbreviated as Word or N-gram based Dynamic Information Retrieval Engine (DIRE). This is written in Java and provides basic indexing, retrieval and storage facilities for documents.

### *B. Owlir Architecture*

It is described as follows:

(a) Information Extraction (IE):- This part is included in its architecture in order to make text documents to semantic web documents and this can be done using IE tools.

***Approach involved*: -** OWLIR uses Aero Text system which is used to extract text of key phrases and elements from free text documents.

**(b) Inference System (IS):** - OWLIR uses metadata information of text to find semantic relations. These relations will decide scope of search and provides effective responses. OWLIR functionality is based on DAML Jess KB where Jess is *Java Expert System Shell.*

DAML Jess KB facilitates reading and interprets DAML+OIL pages and gives reason to users for using that information.

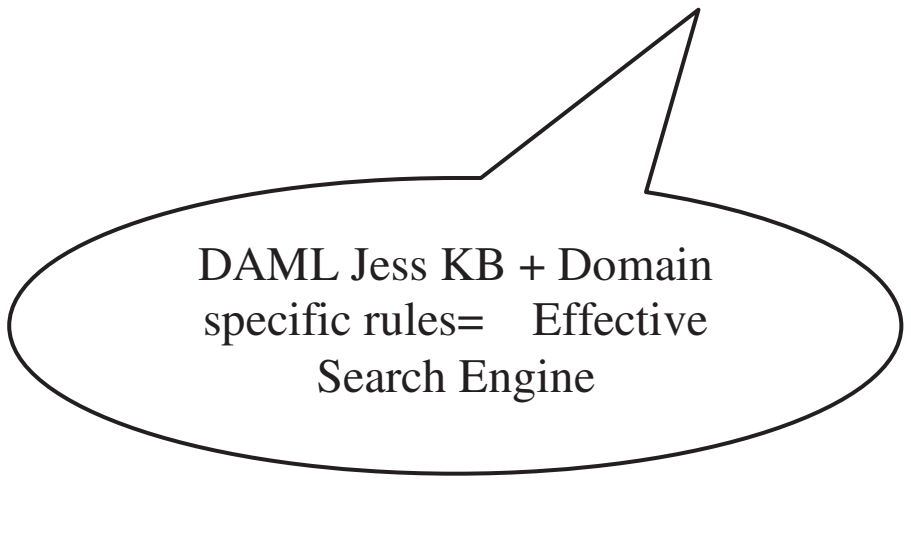

**Fig. 7: "Search Engine"**

### *C. Flow of Information in OWLIR*

Documents are processed by extraction tools like Aero Text. It produces DAML+OIL [6] markup. Then, RDF triples are generated from DAML+OIL pages. Additional RDF triples are extracted from web and forms Inference Engine (IE).

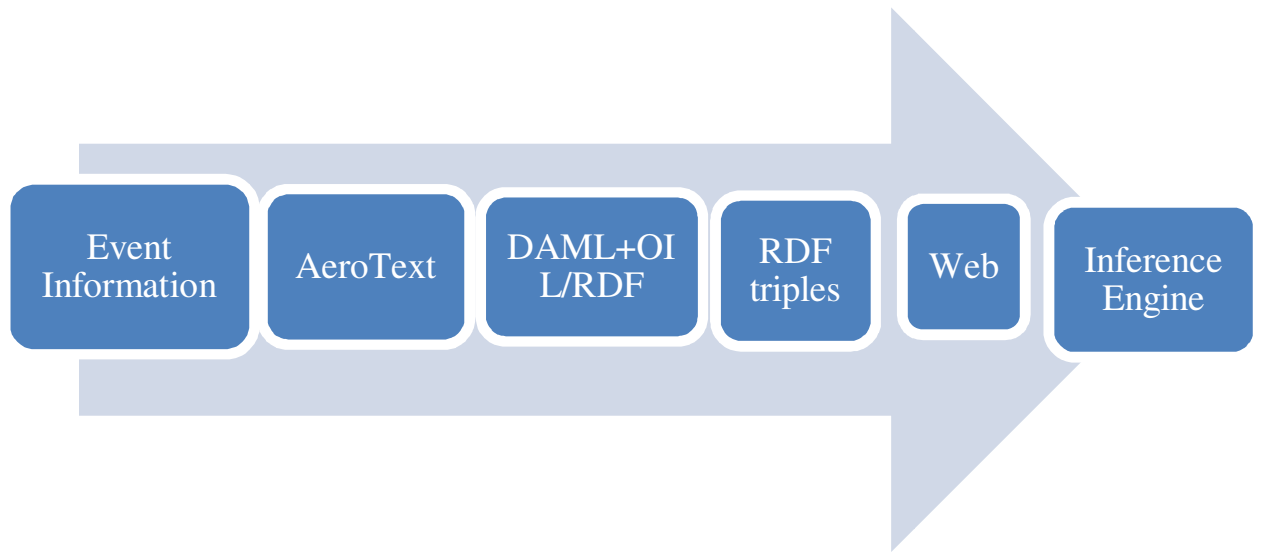

**Fig. 7: "Information flow"**

### *D. Swangler*

This is one of prototype systems and metadata engine. It annotates RDF documents encoded in XML and produces documents that are compatible with Google and other engines. Google treats SWD's as text files. But this creates following two main problems:

(i) XML name space is not valid to search engines like Google.

(ii) Tokenization rules are designed for natural languages.

***Solution*: -** SWANGLING is used to enrich SWD's with extra RDF statements. RDF files are modified and then put on web for Google to discover. When it is discovered, Google indexes the contents using swangler.

### *E. Swoogle*

We have developed prototype search engine called SWOOGLE [7] to facilitate the development of Semantic Web. With the help of Swoogle, we can Access, Explore and Query (AEQ) RDF and OWL documents.

Swoogle is crawler based indexing and retrieval system for Semantic Web. It extracts metadata for each discovered document and gives relationships between documents. Documents are indexed by IR system which uses character N-gram as keyword to find relevant documents.

### *E.1 Analysis*

After we have developed Swoogle, it is found to be analyzed on three activities which are listed below:-

(i) Helps in searching appropriate ontologies.

(ii) Searching Data Instance.

(iii) Characterize Semantic Web.

These are discussed below in details:

**(a)** ***Searching appropriate Ontology*: -** Conventional Search engines failed many times to find required events for particular task. Swoogle helps in finding ontologies as it allows user to query for documents.

**(b) Finding Data *Instance*: -** Swoogle allows user to query SWD's with keywords that uses Classes/Properties.

**(c)** ***Characterizing Semantic Web*:** Collection of data by researchers' leads to characterization of SW. User can answer any question about ontology.

### *Is Swoogle Better than Ontology Repositories?*

Yes, it is better. Ontology repositories like DAML+OIL, Schema Web do not automatically discover documents. They requires user to submit URL's. They store RDF documents rather than solving them and querying them.

### *E.2 Swoogle Architecture*

Four components include in its architecture. They are as follows:

(i) SWD's discovery

(ii) Metadata creation

(iii) Analysis of data

(iv) Interface

All above four components work independently and interact with each other through database. These are detailed below:

***(i) SWD's discovery*:** It discovers Semantic Web Documents and keeps up to data information about objects.

***(ii) Metadata creation*:** It gives SWD cache and generates metadata at both semantic and syntactic level.

***(iii) Data Analysis*:** It uses cache SWD's and metadata to produce analysis with the help of IR analyzer and SWD analyzer.

***(iv) Interface*:** It provides data services to SW community.

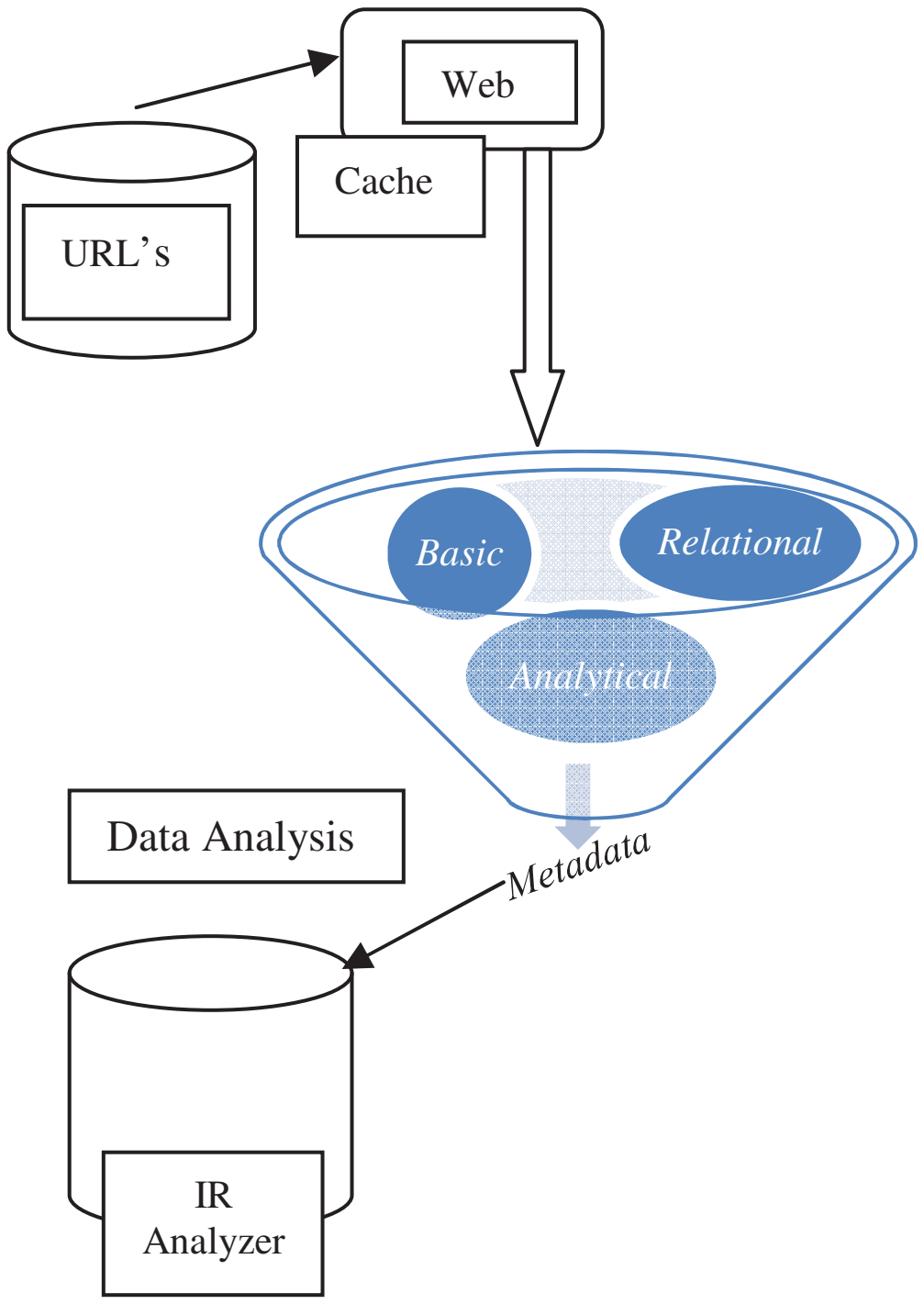

**Fig. 8: "Swoogle Architecture"**

## 6. CONCLUSIONS

The emphasis in the paper is on the concept of Semantic Web and various approaches used for retrieving information from web. Web contains millions of documents and to retrieve relevant information from them, we have gone through various prototypes which act as search engine. Information Retrieval over collection of those documents offers new challenges and opportunities. We have presented framework for integrating search that supports Inference engine. We can use Swangling technique to enrich SWD's to text documents.

Use of OWLIR confirms as semantic markup within documents can be used to improve retrieval performance or not. Swoogle is desired to work with all SWD's. It is better than current web search engines like Google that work with natural languages only.